\documentstyle[epsf,prl,aps]{revtex}
\def\be{\begin{eqnarray}}
\def\ee{\end{eqnarray}}
\def\nn{\nonumber}
\begin{document}
\twocolumn[\hsize\textwidth\columnwidth\hsize\csname@twocolumnfalse\endcsname
\title{\bf Comment on ``Pionic decay of a possible $d'$ dibaryon and
the short-range $NN$ interaction'' }
\author{  A. Samsonov and M. Schepkin}
\address{Institute for Theoretical and Experimental Physics \\
Moscow 117218, Russia}
\maketitle

\begin{abstract}
We comment on calculations of the width of the 
$d'$ resonance within framework of quark shell models. 
\end{abstract}
\vskip1pc]

In a recent paper \cite{O} by I. Obukhovsky, K. Itonaga,
Georg Wagner, A. Buchmann and A. Faessler the decay of the
$d'$ resonance is calculated in a microscopic quark shell model.
This resonance has been suggested to explain peculiarities in the
forward angle cross section of the pionic double charge exchange
(DCX) on nuclei at low energies \cite{BCS}.
Quantum numbers of the resonance, $T(J^P) = 0(0^-)$, imply that
the decay $d' \to $NN$\pi$ is dominated by $s$--waves between
the outgoing particles.  In general the strong interaction 
$d'$NN$\pi$ vertex contains two independent invariant 
amplitudes (see ref.\cite{SZC}). However at low energies
(like e.g. in the decay $d' \to $NN$\pi$) only one amplitude
survives.\footnote{This is easy to see since in that case
the NN--pair is in the $^1S_0$--state ($J^P=0^+$), hence the 
4-body vertex contains 3 spinless ``particles'':
$0^-~(\pi)$, ~~$0^-~(d')$, ~~and $0^+$ (dinucleon).}

As calculated recently in ref.\cite{O}, the decay amplitude 
has to go as ${\bf k}^2$, where ${\bf k}$ is the 3-momentum 
of the outgoing pion in the $d'$ rest frame. 
This result for the $s$--wave decay 
$d' \to $NN$\pi$ is hardly possible to advocate since  
${\bf k}^2$--behaviour would be a feature of a $d$--wave 
(or double $p$--wave within the 3-body system of outgoing 
particles) process.

On the other hand in ref.\cite{SZC} it has been already shown 
that the $d'$ decay amplitude has to be proportional to the
pion 4-momentum  $k_\mu$ since this amplitude has to satisfy Adler 
self-consistency condition. Thus the leading term in the
amplitude is the one containing the time component $k_0$ of the
4-vector $k_\mu$. This result can be corroborated within a model
similar to that used by the authors of ref.\cite{O}. 

In ref.\cite{O}  it is assumed that the $d'$ wave function 
corresponds to the $s^5p^1$ configuration, while the outgoing
$6q$ system, carrying quantum numbers $J^P = 0^+$, $T=1$, 
can be a mixture of a number of configurations.
Let us take into account one of them, namely $s^6$ configuration.
It will be easy to see, that  the result holds true for the other 
configurations as well.  

In the model considered in ref.\cite{O} the NN$\pi$
decay of the $d'$ is due to the presence of 
the $qq\pi$--vertex, giving rise to the transition
\be
(s^5p^1) \to (s^6) + \pi~,
\label{tr}
\ee
followed by the subsequent fall-apart of the outgoing 
$6q$--system into  NN pair in the $^1S_0$ state.

The $qq\pi$--vertex can be written in the pseudo-scalar 
form,  
${\bar \psi} \gamma_5 {\vec \tau} \psi {\vec \pi}$, or
preferably in the pseudo-vector form \cite{TEO}:
\be 
f_{qq\pi}{\bar \psi} \gamma_\mu \gamma_5 {\vec \tau} \psi 
\cdot \partial_\mu {\vec \pi}
\label{piq}
\ee

Thus it is explicitly seen that a vertex
of a pion emission in any elementary act
is proportional to the pion 4-momentum. This does not exclude,
 of course, an extra dependence of the whole amplitude of a
 physics process on the pion momentum.
The coupling constant $f_{qq\pi}$ has the dimension of length,
and can be fixed so as to reproduce the NN$\pi$ coupling 
constant. 

The probability amplitude of the transition (\ref{tr}) is 
proportional to the amplitude of the transition of a quark 
from $p$ to $s$ shell, $q_p \to q_s + \pi$,
accompanied by the pion emission. The latter equals
\be
f_{qq \pi} \int {\bar \psi}_s({\bf r}) \gamma_\mu \gamma_5
{\vec \tau} \psi_p({\bf r}) ~k_\mu ~ {\vec \pi}~
e^{i{\bf kr}} d{\bf r}~, 
\label{ps}
\ee
if the pion is described by a plane wave, and $\psi_s$
and $\psi_s$ are the wave functions (bispinors), describing
quarks on $s$ and $p$ shells, respectively.

Wave function of a fermion in a state with the definite
total angular momentum $j$,
its projection $m$, and parity $P$ reads as:
\be
\psi_{jm}({\bf r}) = \left (f(r) \Omega_{jlm} ({\bf n})\atop 
(-1)^{\frac{1+l-l'}{2}} g(r) \Omega_{jl'm} ({\bf n})\right )~,
\label{psi}
\ee
where $\Omega_{jlm}$ are spherical spinors depending on
${\bf n}={\bf r}/r$, $r=|{\bf r}|$ (see e.g. \cite{BLP}).
$l=j \pm 1/2$, and $l'=2j-l$. 
For a given $j$ the states with $l=j-1/2$ and $l=j+1/2$
have different parity. The spherical spinor $\Omega_{jl'm}$
can be expressed through $\Omega_{jlm}$ as:
\be
\Omega_{jl'm}=i^{l-l'}~
(\mbox{\boldmath$\sigma $}{\bf n})
\Omega_{jlm}~.
\label{Om}
\ee

For a quark on $s_{1/2}$ and $p_{1/2}$ shells the 
corresponding bispinors (\ref{psi}) look particularly simple:
\be
\psi_{s_{1/2}} = \left (f_0(r) \varphi \atop 
g_0(r)(\mbox{\boldmath$\sigma $}{\bf n}) \varphi \right )~,
\label{s12}
\ee
and
\be
\psi_{p_{1/2}} = 
\left (f_1(r)(\mbox{\boldmath$\sigma $}{\bf n}) \chi \atop 
 g_1(r)  \chi  \right )~.
\label{p12}
\ee
Here $\varphi$ and $\chi$ are nonrelativistic 2-component
spinors which do not depend on ${\bf n}$.

It is easy to see that the leading contribution in 
eq.(\ref{ps}) is the one proportional to the total energy 
of the pion, $k_0$.
\footnote{
Another simple way to prove this statement is to consider
3-body vertex of the $d'$ decay into two spinless
``particles'', pion and di-nucleon [NN]. Let $\vec \pi$,
$\vec \Phi$ and $\Psi$ be the operators, annihilating pion,
[NN] and $d'$, respectively. (Here the sign vector stands for
the isospin degrees of freedom).
 The only way to construct the 
Lorentz invariant vertex describing emission of a pion by
the axial current is: 
$({\vec \Phi}^+ \stackrel{\leftrightarrow}{\partial}_{\mu}
 \Psi)~(\partial_{\mu} \vec \pi)$.
The amplitude of the decay $d' \to$[NN]$\pi$ is then 
proportional to the scalar product of the 4-vectors
$k(P_{[NN]}+P_{d'})$, where $P_{d'}=P_{[NN]}+k$.
This scalar product equals $M_{d'}^2-M_{[NN]}^2$; 
for a small energy release, $\Delta E=M_{d'}- M_{[NN]}$
(which is the case in the $d'$ decay), it equals
$2\Delta E \cdot M_{[NN]} \approx 4 M_N k_0$ to a good 
accuracy since $\Delta E \approx k_0$, while extra
terms proportional to powers of ${\bf k}^2$ are very small.}

For $\psi_{s,p}$ given by eqs.(\ref{s12}) and (\ref{p12})
the expression (\ref{ps}) can be presented in a simple form:
\be
f_{qq\pi} (\varphi^+ \chi)\left [-k_0 
\int(g_0^*f_1+f_0^*g_1) e^{i{\bf kr}}  d{\bf r}   \right.
\nn      \\
 + \int (f_0^*f_1 + g_0^*g_1) ({\bf kn})e^{i{\bf kr}} 
 d{\bf r} ~, 
\label{int}
\ee
in which integration over angles is trivial.

Further details depend of course on the potential
used to find solution for $\psi_{s,p}$, and in particular
on the Lorentz structure of the potential. 

For the $qq\pi$ vertex written in the pseudo-scalar form
the amplitude of the  transition (\ref{tr}) is proportional to
\be
(\varphi^+ \chi)
\int(g_0^*f_1-f_0^*g_1) e^{i{\bf kr}}  d{\bf r}  ~.
\label{PSc}
\ee

Thus we conclude that for both types  
of the $qq\pi$ coupling (pseudo-vector and  pseudo-scalar)
the amplitude of the
$s$--wave decay $d' \to$NN$\pi$ does not vanish at 
${\bf k} \to 0$, as distinct from the statement in  
ref.\cite{O}, according to which the amplitude is 
proportional to ${\bf k}^2$. The result of ref.\cite{O}
is therefore appears to be erroneous.

The amplitude of the $s$--wave decay $d' \to$NN$\pi$ is strongly
modified by the NN final state interaction, as had been shown
in our paper from 1993 (see ref.\cite{SZC}), in particular if 
the decay process is due to a point-like interaction responsible
for a $q\bar q$--pair creation.\footnote{
The numerical result for the enhancement caused by the NN FSI
 is of course a model dependent.} This effect is known to 
 give preference to small NN invariant masses, and might be
 a crucial for the experimental searches for the $d'$.
 Thus e.g. by applying a cut in the NN invariant masses one 
 expects an enhancement of the signal-to-background ratio.
 It is this procedure which has been applied to sense 
 the $d'$ contribution in the experiment on double pion
 production, $pp \to pp\pi^-\pi^+$, performed at 
 ITEP (Moscow) \cite{ITEP}.

In ref.\cite{SZC} we considered the simplest case to take
into account NN FSI:\\
1) point like $d'$NN$\pi$ vertex, ~~~and\\
2) Yamagichi wave function for the NN in the continuum.

Decay of the $d'$ into $pp\pi^-$ is also affected by Coulomb
effects (see pion spectra in the decays $d' \to nn\pi^+$
and $d' \to pp\pi^-$ in ref.\cite{SZC}), leading to some 
10 -- 15 \% difference in the decay rates $d' \to nn\pi^+$
and $d' \to pp\pi^-$.

The effects of the NN FSI in the decay  $d' \to pp\pi^-$
can be taken into account (see e.g. \cite{LL}) 
by multiplying the differential probability of the decay by 
\be
F_{pp}(p) = F_C(pa_c)
\label{fsipp}
\ee
\be
\times
\left |1+\frac{\beta +ip}{-a_s^{-1}+\frac{1}{2}r_0p^2-
\frac{2}{a_c}h(pa_c)-ipF_C(pa_c)}\right |^2 
\nn
\ee
where $p \equiv |{\bf p}|$, and ${\bf p}$ is the 3-momentum
of either proton in the $pp$ c.o.m.
Functions $F_C(x)$ and $h(x)$ are given by:
\be
F_C(x)=\frac{\frac{2\pi}{x}}{e^{\frac{2\pi}{x}}-1}
\ee
and
\be
h(x)=\frac{1}{x^2}\sum_{n=1}^\infty 
\frac{1}{n(n^2+x^{-2})} - \gamma + \ln (x)~,
\ee
where $\gamma = 0.577...$ is the Euler constant.
$a_s$ is the $pp$ scattering length, $r_0$ -- effective
range, and $a_c$ is the Bohr radius for the $pp$ subsystem;
$\beta \approx 230$ MeV is the parameter of the Yamaguchi 
potential. Numerical result obtained with eq.(\ref{fsipp})
is only 20--30\% different from the ``exact'' one (see e.g.
recent preprint \cite{Kudr}) obtained from the solution of
the wave equation with ``Coulomb + Yamaguchi potential''
\cite{CY}. 

Thus we see that the dependence of the $d'$NN$\pi$ amplitude
on the pion momentum, $k_\mu$, in combination with the
FSI effects appears to be extremely important for any
spectra in the $d'$ decay. An extra factor ${\bf k}^4$ in
the amplitude squared would have led to even stronger 
enhancement of the high energy side of the pion spectrum
(corresponding to small $pp$ invariant masses).
However this is not so, and $k$--dependence of the
invariant amplitude squared is given by the product
$k_0^2F_{pp}(p)$, where $k_0$ is the total energy of the
pion in the $d'$ rest frame.

Potential models similar to that used in ref.\cite{O}
should lead to a vanishingly small radiative decay rate
$d' \to d \gamma$. This decay mode is of special interest
since $\gamma d$ is the only elementary process where
$d'$ should manifest itself as a Breit--Wigner pole in
$s$--channel. The nature of suppression of the isoscalar
E1 transition $d' \to d\gamma$ is the same as in 
nuclear physics where E1 transitions with $\Delta T = 0$
between nuclei with $N = Z$ are known to be forbidden
(L.Radicati, 1952, see e.g. \cite{LL}). 
The same holds true not only for potential quark models,
but also for any quark cluster model where (effective) 
mass of a quark cluster is proportional to the number 
of quarks in the cluster. An example, demonstrating
this statement was considered in ref.\cite{gamma},
where it was assumed that the $d'$ represents an 
orbitally excited state composed of a diquark
$ud$ $(S=T=0,~~colour=\bar 3)$ with angular momentum
$l=1$ relative to a four-quark cluster
$uudd$ $(S=1,~T=0,~~colour=3)$. The corresponding
E1 amplitude of the transition $d' \to d\gamma$ is then
proportional to $e_2/m_2 - e_4/m_4$, where $e_n$ and $m_n$
are electric charge and mass of the $n$--quark cluster,
respectively. The E1 amplitude vanishes if $m_4 =2m_2$
because $e_4=2e_2 ~ (= 2/3)$. 

For the configuration $s^5p^1$ considered in ref.\cite{O}
sum of the E1 amplitudes corresponding to the two
possible splittings $u~-~uuddd$ and $d~-~uuudd$ also 
vanishes if $m_5=5m_q$. 

Let us note that in the potential model used in ref.\cite{O}
the $d'$ mass appears to be too large, while in the
flux-tube model with anomalously light diquark $ud~(S=T=0)$
the mass of the  $0^-$ isoscalar $6q$--state is rather
close to the one needed for the dibaryon explanation of
the peculiarities in the pionic DCX. Simultaneously the
existence of light diquarks makes ineffective the suppression
of the E1 transitions discussed above. \\

We would like to thank T.Ericson, H.Clement,
B.Loiseau, M.Krivoruchenko, B.Martemyanov, 
A.Buch-mann
and I.Obukhovsky for useful discussions.
This work was supported in part by the Swedish Academy of 
Sciences and the National Research Council (NFR) and by 
INTAS-RFBR Grant 95-605.


\begin{thebibliography}{12}
\bibitem{O} I.Obukhovsky, K.Itonaga,
Georg Wagner, A.Buchmann and A.Faessler, nucl-th/9708056,
Phys. Rev. {\bf C}, {\bf 56}, 3295 (1997).
\bibitem{BCS} R.Bilger, H.Clement, M.Schepkin,
Phys.Rev.Lett., {\bf 71}, 42 (1993).
\bibitem{SZC} M.Schepkin, O.Zaboronsky, H.Clement, Z. Phys. 
{\bf A345}, 407 (1993).
\bibitem{TEO} T.E.O.Ericson and W.Weise, {\sl Pions and
Nuclei}.  Clarendon Press, Oxford, 1988.
\bibitem{BLP} V.B.Berestetskii, E.M.Lifshitz and
L.P.Pitaevskii, {\sl Relativistic Quantum Theory}.
Vol.1, ``{\sl Nauka}'', Moscow, 1968.
\bibitem{ITEP} L.S. Vorobyev et al., JETP Lett., 
{\bf 59}, 75 (1994)
\bibitem{LL} L.D.Landau and E.M.Lifshitz, 
{\sl Quantum Mechanics}. ``{\sl Nauka}'', Moscow, 1963.
\bibitem{Kudr} B.L.Druzhinin, A.E.Kudryavtsev, V.E.Tarasov.
Preprint ITEP 41-96 (1996).
\bibitem{CY} H. van Haeringen, Nucl.Phys. {\bf A 253}, 355
(1975).
\bibitem{gamma} R.Bilger et al., Nucl.Phys.,
{\bf A 596}, 586 (1996).

\end{thebibliography}
\end{document}